\documentclass[12pt, preprint]{aastex}

\usepackage[]{natbib}

\begin{document}
\title{Discrete Intrinsic Redshifts from Quasars to Normal Galaxies}
\author{M.B. Bell\altaffilmark{1}}
\altaffiltext{1}{Herzberg Institute of Astrophysics,
National Research Council of Canada, 100 Sussex Drive, Ottawa,
ON, Canada K1A 0R6}

\begin{abstract}

It is pointed out that the discrete velocities found by Tifft in galaxies are harmonically related to the discrete intrinsic redshifts found in quasars. All are harmonically related to the constant 0.062$\pm0.001$, and this is the fourth independent analysis in which the redshift increment 0.062 has been shown to be significant. It is concluded that there is a quantized component in the redshift of both quasars and galaxies that has a common origin and is unlikely to be Doppler-related. 
 
\end{abstract}


\section{Introduction}
 
Over the last thirty years much evidence has been presented suggesting that the redshifts of both quasars \citep{bur68,bur90,kar71,kar77,bur01,bel02a,bel02b,bel02c} and galaxies \citep{tif96,tif97} may contain a quantized component. Here, the available evidence on discrete redshifts and velocities in quasars and galaxies is re-examined in light of recent new results.

\section{Discrete Redshift and Velocity Data in Quasars and Galaxies}

\subsection{Discrete Redshifts in Quasars with z $< 0.6$}

Over thirty years ago \citet{bur68} reported that the redshift z = 0.061 was observed frequently in quasar redshifts. \citet{bur90} found that quasar redshifts less than 0.6 appeared to be clumped at intervals close to z = 0.06. Thus these quasars and QSOs found in early surveys, presumably mostly radio sources detected above the relatively high detection limit of the time, appeared to have redshifts that favored discrete intervals that were pure harmonics of 0.061 $\pm$0.002. For these discrete intervals to be visible in the data would require that the cosmological component of their redshifts be less than z$_{\rm c} \sim 0.02$.

\subsection{Discrete Redshifts in Quasars with z$ < 2$}

Recently it has been pointed out \citep{bel02c} that all quasar redshifts may contain a discrete intrinsic component that is a harmonic of 0.062. These harmonics need not be spaced consecutively. The redshift values predicted by \citet{bel02c} had already been found in early redshift distributions although the clumping had been claimed to be periodic in log(1+z) \citep{kar71,kar77}. It was not noticed that the peaks lay at harmonics of z = 0.062, although this quickly became apparent after all Doppler components were removed from the measured redshifts of the QSOs near NGC 1068 \citep{bel02c}. For the peaks in the redshift distribution to be visible, most of the objects must be nearby (z$_{\rm c} < 0.01$). If they are not, the peaks will be smeared out by the cosmological redshift component. Therefore it is assumed that any quantization will be most apparent in the early surveys \citep{kar77} where the detection limit was poor and only the closest objects would be detected. Adding additional sources found in later, more sensitive surveys, would be expected to smear out any previous intrinsic quantization because of the larger cosmological redshift component present. Similarly, it has been shown recently \citep{bel02c} that objects ejected in random directions from a parent galaxy can also have sufficiently high l-o-s ejection velocity components to completely smear out the predicted quantized intrinsic component. 

Recently \citet{haw02} have searched for the log(1+z) periodicity in 1647 QSO-galaxy pairs found in the 2dF redshift survey data. No periodicity was detected, as would be expected from the above arguments.
What the 2dF analysis does appear to show, however, is that if the redshifts are purely cosmological, they do not appear to contain a periodicity in log(1+z). 

\citet{bel02c} found evidence that so far all intrinsic redshifts in quasars can be defined by the relation:

\begin{equation}
z_{i}\:= 0.062[10N - M_{N}]
\end{equation}

where $N$ = 1,2,3.... The different M$_{N}$ relations for $N$ = 1, 2, and 3 are given in Table 1. 

Equation 1 actually represents a series of equations, one for each N-value. The first three of these are:
\begin{equation}
z_{i[N=1]}\:= 0.062[10 - n]\;_{n = 0,1,2...9}
\end{equation}

\begin{equation}
z_{i[N=2]}\:= 0.062\left[20 - \frac{n(n + 1)}{2}\right]\;_{n = 0,1,2,3,4,5}
\end{equation}

\begin{equation}
z_{i[N=3]}\:= 0.062\left\{30 - \left( \frac{n(n+1)}{2} \right) \left(\frac{\frac{n(n+1)}{2} + 1}{2}\right) \right\}_{n = 0,1,2,3}
\end{equation}

Intrinsic redshift values defined by these equations are listed in Table 2, cols 2, 3, and 4. The minimum intrinsic redshift for each $N$-value has been highlighted. Note that the minimum intrinsic redshift corresponds to the maximum value that $n$ can have before the intrinsic redshift becomes a blueshift. Also included in the table are values for $N$ = 4, 5, and 6, however, as discussed by \citet{bel02c}, it is not yet known if these states are allowed. So far the source of the intrinsic redshifts has not been identified, other than to suggest that it may be gravity related. This is based on the fact that the maximum redshifts in each $N$-state are separated by z = 0.62. It has been shown that the redshift z = 0.62 is the maximum gravitational redshift that can be obtained in a perfect fluid sphere if appropriate conditions on the equation of state and on stability are applied \citep[and references therein]{bur67}.

\begin{deluxetable}{ccccc}
\tabletypesize{\scriptsize}
\tablecaption{Parameters associated with different values of $N$ in Eqn. 1 \label{tbl-1}}
\tablewidth{0pt}
\tablehead{
\colhead{$N$} & \colhead{$M_{N}$} &  \colhead{n} & \colhead{Data Source} & \colhead{Reference} 
}
\startdata
1 &  n  & 0,1,2,3...9  & quasars (z $< 0.6$) & \citet{bur90} \\
2 & n(n+1)/2  &  0,1,2,3,4,5. & QSOs near NGC 1068 & \citet{bel02c} \\
3 & [p(p+1)/2]\tablenotemark{a} & 0,1,2,3. & extrapolated from $N$=1 and 2. & \citet{bel02c} \\
\enddata 
\tablenotetext{a}{p = n(n+1)/2}
\end{deluxetable}

\begin{deluxetable}{ccccccc}
\tabletypesize{\scriptsize}
\tablecaption{Acceptable Values of Intrinsic Redshift from Eqn 1. \label{tbl-2}}
\tablewidth{0pt}
\tablehead{
\colhead{n} & \colhead{z$_{\rm i}[N$ = 1]} & \colhead{z$_{\rm i}[N$ = 2]} & \colhead{z$_{\rm i}[N$ = 3]} & \colhead{z$_{\rm i}[N$ = 4]\tablenotemark{a}} & \colhead{z$_{\rm i}[N$ = 5]\tablenotemark{a}} & \colhead{z$_{\rm i}[N$ = 6]\tablenotemark{a}} 
}
\startdata
0 & 0.620   &  1.240 &  1.86 & 2.48     & 3.100 & 3.720  \\
1 & 0.558   & 1.178   & 1.798 & 2.418   & ${\bf 3.038}$ & ${\bf 3.658}$  \\
2 & 0.496   & 1.054   & 1.488 & ${\bf 1.178}$   &  ---  & --- \\
3 & 0.434   & 0.868  &  ${\bf 0.558}$   &  ---    &   ---  & --- \\
4 & 0.372   & 0.620  &  ---   &  ---    &   ---  & ---    \\
5 & 0.310   & ${\bf 0.310}$   &  ---   &  ---    &   ---  & ---   \\
6 & 0.248   & ---   &  ---   &  ---    &   ---  & ---   \\
7 & 0.186   & ---  &  ---   &  ---    &   ---  & ---    \\
8 & 0.124   & ---  &  ---   &  ---    &   ---  & ---    \\
9 & ${\bf 0.062}$   & ---  &  ---   &  ---    &   ---  & ---   \\

\enddata 
\tablenotetext{a}{may not be allowed}
\end{deluxetable}

\subsection{Discrete Velocities in Galaxies}

\citet{tif96,tif97} has found families of discrete velocities in normal galaxies and has been able to fit them to a model in which they are all harmonically related to "c", the speed of light. In his model \citep{tif96}, the discrete components detected are octave-spaced sub-harmonics of c (and related harmonics) and the redshift is assumed to arise from time-dependent decay from an origin at the Planck scale. His periods are given by the relation,

\begin{equation}
P = c2^{-\frac{9D+T}{9}}
\end{equation}
where D is the number of doublings and T defines the different families. There can be many different families of periods, but the four most dominant ones found by \citet{tif97} are listed here in Table 3. The most common or basic of these make up what is referred to by \citet{tif97} as the T = 0 family of discrete values found in common spiral galaxies. Eqn 5 predicts discrete velocities that are very close to those actually measured in galaxies. For example, where velocity periods near 36 and 72 km s$^{-1}$ were measured, corresponding values of 36.5958 and 73.1916 km s$^{-1}$ are predicted by the above equation. Since the fit is very close, the values given by this equation for the dominant periods are used here to identify each family of periods. It should be kept in mind, however, that the measured values are smaller by a few percent.


\begin{deluxetable}{ccccc}
\tabletypesize{\scriptsize}
\tablecaption{Common Velocity Periods in km s$^{-1}$ found by Tifft\tablenotemark{a} \label{tbl-3}}
\tablewidth{0pt}
\tablehead{
\colhead{D/T\tablenotemark{b}} & \colhead{7} &  \colhead{6} & \colhead{1} & \colhead{0} 
}
\startdata
17 & ---     & ---      & ---     &  2.2872 \\
16 & 2.6681  & 2.8817   & 4.2354  &  4.5745 \\
15 & 5.3363  & 5.7635   & ---     &  9.1490 \\
14 & 10.6725 & 11.527   & 16.9416 &  18.2979 \\
13 & ---     &  ---     & ---     &  36.5958 \\
12 & ---     &  46.1078 & ---     &  73.1916 \\
11 & ---     &  92.2157 & ---     &  146.3833 \\
10 & ---     &  184.4313 & ---    &  --- \\
\enddata 
\tablenotetext{a}{as given in Table 1 of \citet{tif97}}
\tablenotetext{b}{see text for a definition of D and T}
\end{deluxetable}

\begin{deluxetable}{cccccccccc}
\tabletypesize{\scriptsize}
\tablecaption{Octave-Spaced Sub-Harmonics of Minimum z$_{\rm i}$ Values and related Velocities in km s$^{-1}$. \label{tbl-4}}
\tablewidth{0pt}
\tablehead{
\colhead{m} & \colhead{z$_{\rm i}$[$N$ = 1]} & \colhead{Vel[$N$=1]\tablenotemark{a}} & \colhead{Vel[T=0]} &  \colhead{z$_{\rm i}$[$N$=2]} & \colhead{Vel[$N$=2]\tablenotemark{a}} & \colhead{Vel[T=6]} & \colhead{ z$_{\rm i}$[$N$=3]}  & \colhead{Vel[$N$=3]\tablenotemark{a}} & \colhead{Vel[T=7]}
}
\startdata
0 & 0.062  &  18012  & ---  & 0.31 & 79040.2 & --- & 0.558 & 124851.7  &  --- \\
1 & 0.031 & 9150     &  --- &  0.155 &  42903.6 & --- & 0.279 & 72318.5  & --- \\
2 & 0.0155 & 4610.7  & ---  & 0.0775 & 22336.1 & ---  & 0.140 & 38928.7 & --- \\
3 & 0.00775 & 2314.3 & ---  & 0.0388  & 11392.0  & --- & 0.0698 & 20182.9 & ---  \\
4 & 0.00387 & 1157.9 & ---  & 0.0194   & 5752.2 & --- &  0.0349 & 10273.0 & --- \\
5 & 0.0019 & 580.13  & ---  & 0.00969   & 2890.2 & ---  & 0.0174 & 5182.1 & --- \\
6 & 0.000969 & 290.36 & ---  & 0.00484    & 1448.59 & --- & 0.00872 & 2602.4 & --- \\
7 & 0.000484 & ${\bf 145.15}$ & ${\bf 146.38}$ & 0.00242   & 725.157 & --- & 0.00436 & 1304.1 & --- \\
8 & 0.000242 & ${\bf 72.54}$  & ${\bf 73.19}$  & 0.00121   & 362.798 & ---  & 0.00218 & 652.7 & ---  \\
9 & 0.000121 & ${\bf 36.27}$  & ${\bf 36.60}$  & 0.000605  & ${\bf 181.32}$  & ${\bf 184.4313}$ & 0.00109  & 326.5 &  ---  \\
10 & 0.0000605 & ${\bf 18.15}$ & ${\bf 18.30}$ &  0.000303 &  ${\bf 90.733}$  & ${\bf 92.2157}$ & 0.000545 & 163.3 & ---   \\
11 & 0.0000303 & ${\bf 9.054}$ & ${\bf 9.15}$  &  0.000151  & ${\bf 45.373}$  & ${\bf 46.1078}$ & 0.000272 & 81.67 & ---   \\
12 & 0.0000151 & ${\bf 4.54}$  & ${\bf 4.57}$  &  0.0000757 & 22.687  & --- & 0.000136 & 40.84 & ---  \\
13 & 0.00000757 & ${\bf 2.27}$ & ${\bf 2.287}$ & 0.0000378  & ${\bf 11.332}$  & ${\bf 11.527}$ & 0.0000681 & 20.42 & ---  \\
14 & ---       & ---   & ---  & 0.0000189   &  ${\bf 5.666}$  & ${\bf 5.7635}$ & 0.0000341 & ${\bf 10.21}$ & ${\bf 10.67}$  \\
15 & ---       & ---   & ---  & 0.00000946  & ${\bf 2.836}$  & ${\bf 2.8817}$ & 0.0000170 & ${\bf 5.105}$ & ${\bf 5.336}$  \\
16 & ---       & ---   & ---  & ---         &  --- & ---     & 0.00000851 & ${\bf 2.553}$ & ${\bf 2.668}$ \\
\enddata 
\tablenotetext{a}{ corrected for realtivistic effect}
\end{deluxetable}

\section{Discussion}

In Table 4, octave-spaced sub-harmonics of the first three minimum redshift values highlighted in Table 2 (0.062, 0.310, 0.558) are listed in columns 2, 5, and 8. These have been converted to velocities with relativistic effects included in columns 3, 6, and 9 respectively. The discrete velocities given by Tifft's model (Table 3), are also listed in Table 4, columns 4, 7, and 10 respectively for periods defined by T-values of 0, 6 and 7. These results show that the velocities predicted by Tifft's model for these three dominant T-states are very close to the octave-based sub-harmonics of the minimum redshifts found for the $N$ = 1, 2 and 3 quasar intrinsic redshift states. The latter are again smaller by a few percent and agree more closely with Tifft's measured discrete velocities. Thus Tifft's model appears to predict values a few percent higher than both his measured values and the values predicted here from the quasar minimum discrete redshifts.

Fig. 1 shows how the discrete velocities predicted by Tifft's model for his most dominant states are tied directly to quasar redshifts in each of the first three quasar $N$-states. The sub-harmonics of the minimum value, 0.062, in the $N$ = 1 state correspond to the periods found in common spirals, and as mentioned above, are referred to by \citet{tif97} as the most common or basic family of periods. Similarly, octave-based sub-harmonics of the minimum values 0.310 and 0.558 (which are both harmonics of 0.062) in the $N$ = 2 and $N$ = 3 quasar states are tied to Tifft's other two dominant states (T = 6 and 7). \em Thus the constant 0.062$\pm0.001$ appears to be a number that is not only a significant one in quasar redshifts, it also appears to be related to discrete velocities in galaxies. \em This is an amazing result, and it represents the fourth independent instance in which the number 0.062 has been found to be significant.

Although a fourth family of periods, albeit with only 2 members, was identified by \citet{tif97}, it does not appear to be harmonically related to either quasar $N$-states $N$ = 4 or $N$ = 5. Although it can be fitted to the $N$ = 6 state, for the moment it has been assumed that there is little evidence for $N$-states above $N$ = 3 as concluded earlier \citep{bel02c}. 

\begin{figure}
\hspace{-1.0cm}
\vspace{-2.0cm}
\epsscale{1.0}
\plotone{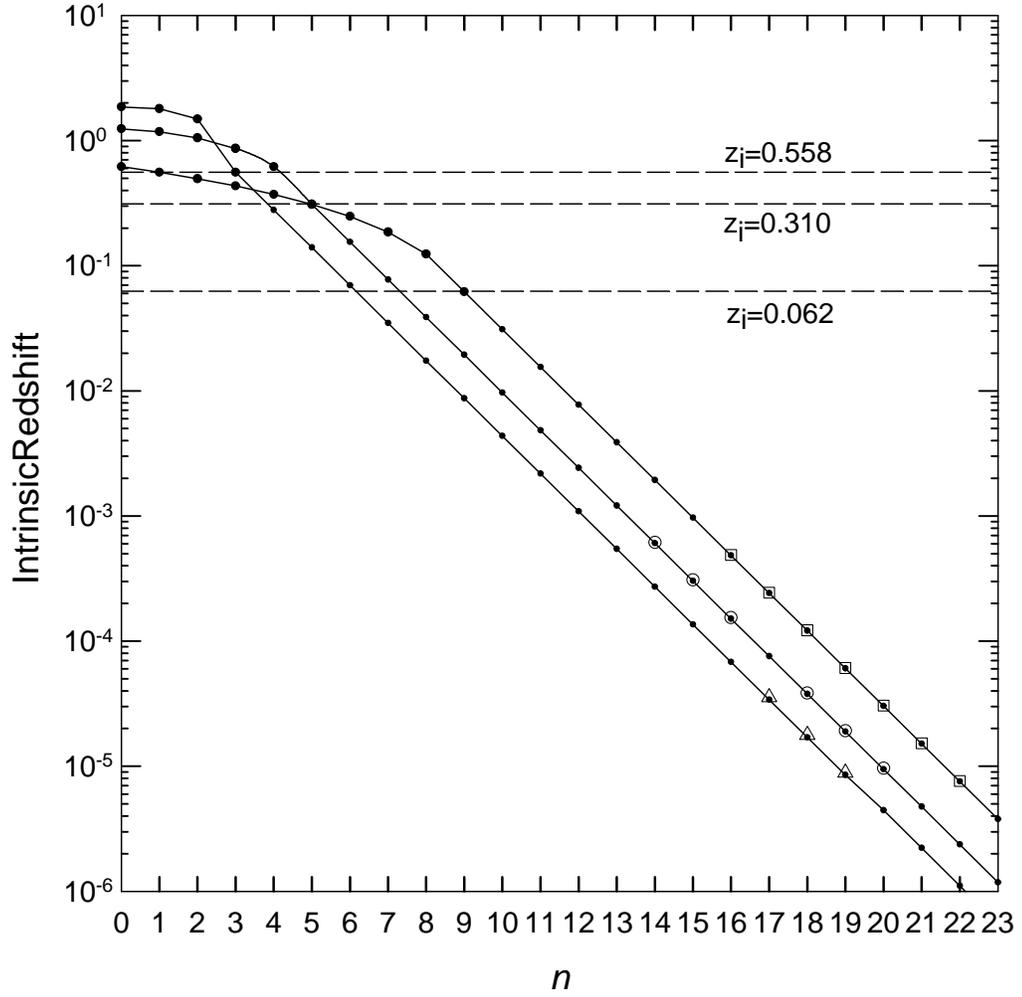}
\caption{\scriptsize{Intrinsic redshifts for all three $N$-values plotted on a logarithmic scale vs $n$. Filled circles above 0.062 represent data found for quasars, open squares are data for T=0 velocities, open circles for T=6 velocities, and open triangles for T=7 velocities. Dashed lines represent the minimum acceptable intrinsic redshifts in the first three $N$-states. \label{fig1}}}
\end{figure}

\begin{figure}
\hspace{-1.0cm}
\vspace{-1.0cm}
\epsscale{1.0}
\plotone{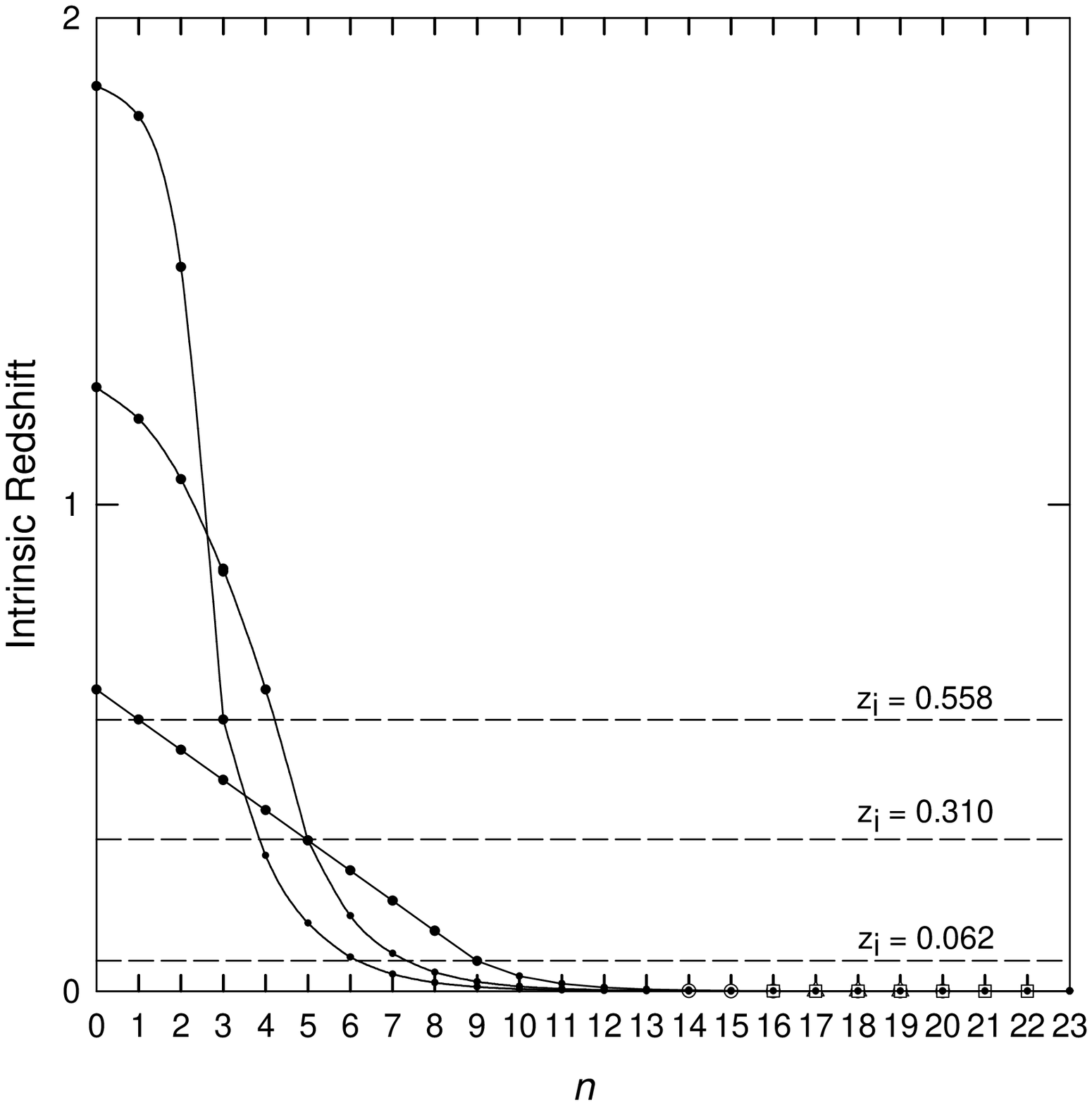}
\caption{\scriptsize{Same as Fig. 1 with intrinsic redshifts for the first three $N$-values plotted on a linear scale vs $n$. \label{fig2}}}
\end{figure}

In Fig. 2, the intrinsic redshift values for the first three $N$-states of equation 1 are plotted on a linear scale. Both Figs. 1 and 2 show clearly that the discrete intrinsic redshifts decrease smoothly and continuously from quasars to galaxies. This smooth continuity from quasars to galaxies, together with the fact that 0.062 can be tied to both types of objects, suggests that all intrinsic redshifts have a common origin. The discrete redshifts are assumed to be intrinsic simply because all known Doppler components are assumed to have been accounted for. 

It is also important to realize that although the intrinsic redshifts may range from zero in galaxies to at least z = 1.86 in quasars, they have been found in objects with cosmological redshift components that correspond to velocities no higher than a few times 10$^{3}$ km s$^{-1}$. Thus there is no need to make a cosmology-dependent correction to obtain accurate harmonic-related values from eqn. 1.

Tifft has also claimed \citep{tif02a} that his model explaining the discrete velocities in galaxies can explain the discrete redshifts reported in quasars. However, his model, referred to hereafter as the Lehto-Tifft model \citep{tif02b}, assumes that the entire redshift is quantized, unlike the evolutionary model proposed by \citet{bel02b} in which the intrinsic component is superimposed on top of the Hubble flow. This represents a major difference between the two models and means that they are incompatible. This difference has been used by \citet{bel02d} to rule out the Lehto-Tifft model. The good fit between the Lehto-Tifft model and the discrete velocities found in galaxies may therefore have been fortuitous, occurring simply because the number 0.062, claimed here to be linked to all quasar and galaxy redshifts, is also an octave-spaced sub-harmonic of the speed of light which is the basis of the Lehto-Tifft model. In this regard it is also noted that z$_{\rm g}$ = 0.62, the maximum gravitational redshift obtainable in a fluid sphere, is also related to the speed of light. Therefore the suggestion that gravity may play a role in producing intrinsic redshifts may also have resulted from a fortuitous agreement between numbers. Furthermore, if gravity does play a role it may already be possible to say that it is unlikely to be a simple one related only to the mass of normal matter present, since the intrinsic redshift also appears to be inversely related to the age of the object, where age is defined as the elapsed time since the object was born \citep{bel02a,bel02b,bel02c}. If mass were the only important factor, this would seem to imply that it decreases with time.

\section{Conclusions}

It has been shown that the discrete intrinsic redshifts found in quasars \citep{bur90,bel02c} are tied directly to the discrete velocities found to be present in normal galaxies by \citet{tif96,tif97}. All are harmonically related to z = 0.062$\pm0.001$. This is the fourth independent instance in which the redshift z = 0.062 has been found to be significant. It is concluded that the discrete increments found in galaxies are unlikely to be Doppler-related. These results are interpreted as a solid confirmation of the reality of discrete intrinsic redshifts and of the significance of the redshift increment 0.062. 

\acknowledgements

I wish to thank Drs. D. McDiarmid and J.K.G. Watson for helpful comments when this manuscript was being prepared. I also want to thank Simon Comeau for helpful comments and assistance with the data analysis.

\end{document}